\documentclass[a4paper,twocolumn,amsmath,amssymb,aps,preprintnumbers,superscriptaddress,showpacs,longbibliography]{revtex4-1}

\pdfoutput=1 

\usepackage[latin9]{inputenc}
\usepackage{textcomp}
\usepackage{amsmath}
\usepackage{amssymb}
\usepackage{graphicx}

\usepackage{dcolumn}
\usepackage{bm}
\usepackage{epsfig}
\usepackage[usenames]{color}
\usepackage{xcolor}

\definecolor{pinegreen}{rgb}{0.0, 0.47, 0.44}
\definecolor{persiangreen}{rgb}{0.0, 0.65, 0.58}
\definecolor{pakistangreen}{rgb}{0.0, 0.4, 0.0}
\definecolor{mossgreen}{rgb}{0.68, 0.87, 0.68}

\newif\ifcom
\newif\ifdel

\comtrue                    
\deltrue                      

\begin{document}
\title{Ferroelectricity and resistive switching in BaTiO$_3$ thin films with liquid electrolyte top contact for bioelectronic devices}

\author{Maximilian~T.~Becker}
\email{mtb57@cam.ac.uk}
\affiliation{%
Department of Materials Science \& Metallurgy, University of Cambridge, 27 Charles Babbage Road,
Cambridge CB3 OFS, United Kingdom}

\author{Poppy~Oldroyd}
\affiliation{%
Department of Engineering, University of Cambridge, 9 JJ Thomson Avenue,
Cambridge CB3 OFA, United Kingdom}

\author{Nives~Strkalj}
\affiliation{%
Department of Materials Science \& Metallurgy, University of Cambridge, 27 Charles Babbage Road,
Cambridge CB3 OFS, United Kingdom}

\author{Moritz~L.~M\"uller}
\affiliation{%
Department of Materials Science \&  Metallurgy, University of Cambridge, 27 Charles Babbage Road,
Cambridge CB3 OFS, United Kingdom}

\author{George~G.~Malliaras}
\affiliation{%
Department of Engineering, University of Cambridge, 9 JJ Thomson Avenue,
Cambridge CB3 OFA, United Kingdom}

\author{Judith~L.~MacManus-Driscoll}
\affiliation{%
Department of Materials Science \&  Metallurgy, University of Cambridge, 27 Charles Babbage Road,
Cambridge CB3 OFS, United Kingdom}

\date{\today}

\begin{abstract} 
We investigate ferroelectric- and resistive switching behavior in 18-nm-thick epitaxial BaTiO$_3$ (BTO) films in a model electrolyte--ferroelectric--semiconductor (EFS) configuration.
The system is explored for its potential as a ferroelectric microelectrode in bioelectronics.
The BTO films are grown by pulsed laser deposition (PLD) on semiconducting Nb-doped (0.5 wt\%) SrTiO$_{3}$ (Nb:STO) single crystal substrates.
The ferroelectric properties of the bare BTO films are demonstrated by piezoresponse force microscopy (PFM) measurements.
Cyclic voltammetry (CV) measurements in EFS configuration, with phosphate buffered saline (PBS) acting as the liquid electrolyte top contact, indicate characteristic ferroelectric switching peaks in the bipolar current-voltage loop. 
The ferroelectric nature of the observed switching peaks is confirmed by analyzing the current response of the EFS devices to unipolar voltage signals.
Moreover, electrochemical impedance spectroscopy (EIS) measurements indicate bipolar resisitive switching behavior of the EFS devices, which is controlled by the remanent polarization state of the BTO layer.
Our results represent a constitutive step towards the realization of neuroprosthetic implants and hybrid neurocomputational systems based on ferroelectric microelectrodes.
\end{abstract} 
\maketitle

\section{Introduction} 
\label{sec:Introduction}
Bioelectronic interfacing of nerve cells with microelectrode arrays (MEAs) is the basis for neural stimulation and recording in neuroprosthetic devices and hybrid neurocomputational systems.
To provide the electro--neural interface in MEA technology, two main types of materials have been utilized up to now: (i) conductive materials and (ii) insulating dielectric materials.

(i) The traditional approach is based on conductive  microelectrodes, which are in direct contact with neural tissue, as demonstrated by Thomas \textit{et al.} in their seminal work in 1972 \cite{Thomas1972}.
Since then, bioelectronic interfacing with conductive microelectrodes has been intensively studied, resulting in a widespread utilization for \textit{in-vitro} and \textit{in-vivo} applications \cite{Cogan08,Frey10,Gkoupidenis17,Rivnay18}.
However, it is difficult to avoid toxic electrochemical effects (Faradaic currents) at the conductive microelectrode/electrolyte interface, which may result in corrosion of electrodes and cell or tissue damage~\cite{Merrill05} and detrimentally affects long-term stability of neural implants \cite{Oldroyd21}.

(ii) As a different approach, bioelectronic interfacing based on insulating dielectric materials has been introduced in the 1990s, by the demonstration of recording of neural activity with an insulated-gate field-effect transistor \cite{Fromherz1991} and capacitive neural stimulation with an insulated microelectrode on a silicon chip \cite{Fromherz1995}.
State-of-the-art MEAs based on this approach have been fabricated in complementary metal-oxide-semiconductor (CMOS) technology and comprise of an array of capacitive stimulation spots interleaved with a transistor array for bidirectional neuroelectronic interfacing \cite{Bertotti14}.
In order to achieve a purely capacitive coupling between the neuronal signals and the stimulation and recording sites, the entire chip surface of a capacitive CMOS-MEA is covered by a thin ($<$ 30 nm) insulating dielectric layer.
The advantage of the dielectric insulation layer is, that toxic electrochemical reactions are suppressed.
Moreover, capacitive CMOS-MEAs enable simultaneous recording and stimulation across the entire active area, which is beneficial for electrical imaging of neuronal activity \cite{Zeck17,Corna21}.
However, the charge injection capacity (CIC) of capacitive microelectrodes is about three orders of magnitude below state-of-the-art conductive microelectrodes \cite{Becker21}, resulting in a dramatically less stimulation efficacy.
As a consequence, neuroprosthetic devices are typically based on conductive microelectrodes.
For example, state-of-the-art retinal implants to restore vision in blind patiens are based on conductive microelectrodes consisting of Pt \cite{Zhou13} or sputtered IrOx \cite{Mathieson12,Haas20}.

Very recently, a new concept of bioelectronic interfacing -- ferroelectric microelectrodes -- has been introduced in Ref.~\onlinecite{Becker21}, to overcome the problem of low CIC provided by insulated microelectrodes.
Within this concept, it has been proposed to utilize microelectrodes that are coated with an insulating ferroelectric layer instead of a dielectric layer and it has been demonstrated theoretically, that the ferroelectric polarization current contributes to the extracellular stimulation current.
Depending on the remanent polarization $P_{\rm r}$ of the utilized ferroelectric, the CIC of ferroelectric microelectrodes can be increased by up to two orders of magnitude as compared to the commonly used capacitive stimulation \cite{Becker21}, i.e.
\begin{eqnarray} 
\rm{CIC}&=&C_sV_0 + 2P_{\rm r}  \quad  ,
\label{eq:CIC_FE}
\end{eqnarray} 
with $C_s$ being the specific capacitance of the microelectrode and $V_0 $ being the amplitude of the applied voltage signal $V(t)$.
For the successful integration of ferroelectric microelectrodes into bioelectronic devices (e.g. CMOS-MEAs), it is important to mimic the physiological conditions. In order to do that, the utilized ferroelectric should be operated with a liquid electrolyte top contact.
A possible choice for the liquid electrolyte top contact is phosphate-buffered saline (PBS) solution, which is different compared to a conventional metallic (solid) top contact \cite{Gerischer1990,Wallrapp06}.
Moreover, the ferroelectric layer is also required to be of thickness range below~30 nm, such that its specific capacitance is large enough to enable efficient capacitive coupling for simultaneous transistor recording of neural activity.
As an alternative to extracellular recording with field-effect transistors -- which corresponds to a three-terminal device configuration -- the question is raised if recording with ferroelectric microelectrodes can also be achieved in simpler two-terminal electrolyte--ferroelectric--semiconductor (EFS) device configuration.
Theoretically, recording of neural activity with EFS devices can be realized if the devices exhibit resistive switching behavior, i.e, a change in their resistance state due to extracellular voltages.
So far, polarization-modulated resistive switching behavior has been demonstrated in metal-ferroelectric-metal (MFM) devices such as ferroelectric tunnel junctions (FTJs) \cite{Kohlstedt05,Zhuravlev05} and ferroelectric diodes \cite{Blom1994}.
%
%
Although there are reports of polarization reversal in ferroelectric layers with an electrolyte top contact, the utilized ferroelectric layers were either in the thickness range above 100~nm, or the top contact was a non-physiological (solid) electrolyte \cite{Small12,Ferris13,Fabiano14,Toss17}.
Thus, ferroelectricity and resisitve switching in a thin ferroelectric film ($<$ 30 nm) with a liquid electrolyte top contact, has not yet been demonstrated.
In this work, we experimentally demonstrate ferroelectric polarization reversal and resistive switching in 18-nm-thick BaTiO$_3$ (BTO) thin films in a model EFS configuration, a system of great potential interest within the field of bioelectronics, e.g. for future applications of ferroelectric microelectrodes.
Here, we address electrical stimulation and recording in neuroprosthetic implants based on EFS microelectrodes. We also address the feasiability of coupling artificial neural networks with biological networks in hybrid neurocomputational systems based on ferroelectric microelectrodes.

\section{Experimental} 
\label{sec:Experimental}

A stoichiometric BaTiO$_3$ target was used to grow epitaxial single crystalline films on  (001)-oriented Nb-doped (0.5 wt\%) SrTiO$_{3}$ (Nb:STO) single crystal substrates (5 $\times$ 5 mm) by means of pulsed laser deposition (PLD).
Prior to PLD deposition, the substrates were heated up in vacuum to $950\,^{\circ}$C for 1 h to promote formation of step terraces.
For the deposition of 18-nm-thick BTO films, the temperature $T$ was fixed at $650\,^{\circ}$C with an oxygen partial pressure of $p_{\rm O_2}=5\,$mTorr~\cite{Strkalj19}.
We used a KrF excimer laser with laser fluence $E_{\text{L}}$ = 1.2 J/cm$^2$  and laser repetition rate $f_{\text{L}}$ = 2\,Hz.
After deposition, $p_{\rm O_2}$ was increased to 300\,Torr and kept at $T$ = $750\,^{\circ}$C for 15 min, before cooling down to room temperature with a rate of 5\,K/min.

The microstructure of the films was investigated by x-ray diffraction (XRD), x-ray reflectivity (XRR) and atomic force microscopy (AFM) to obtain information on crystallinity, thickness and roughness.
Piezoresponse force microscopy (PFM)  measurements were performed with Pt-coated tips and the PFM contrast was recorded in Cartesian coordinates to remove background signals \cite{Jungk07}.

\begin{figure}[t] 
\includegraphics[width=1\columnwidth]{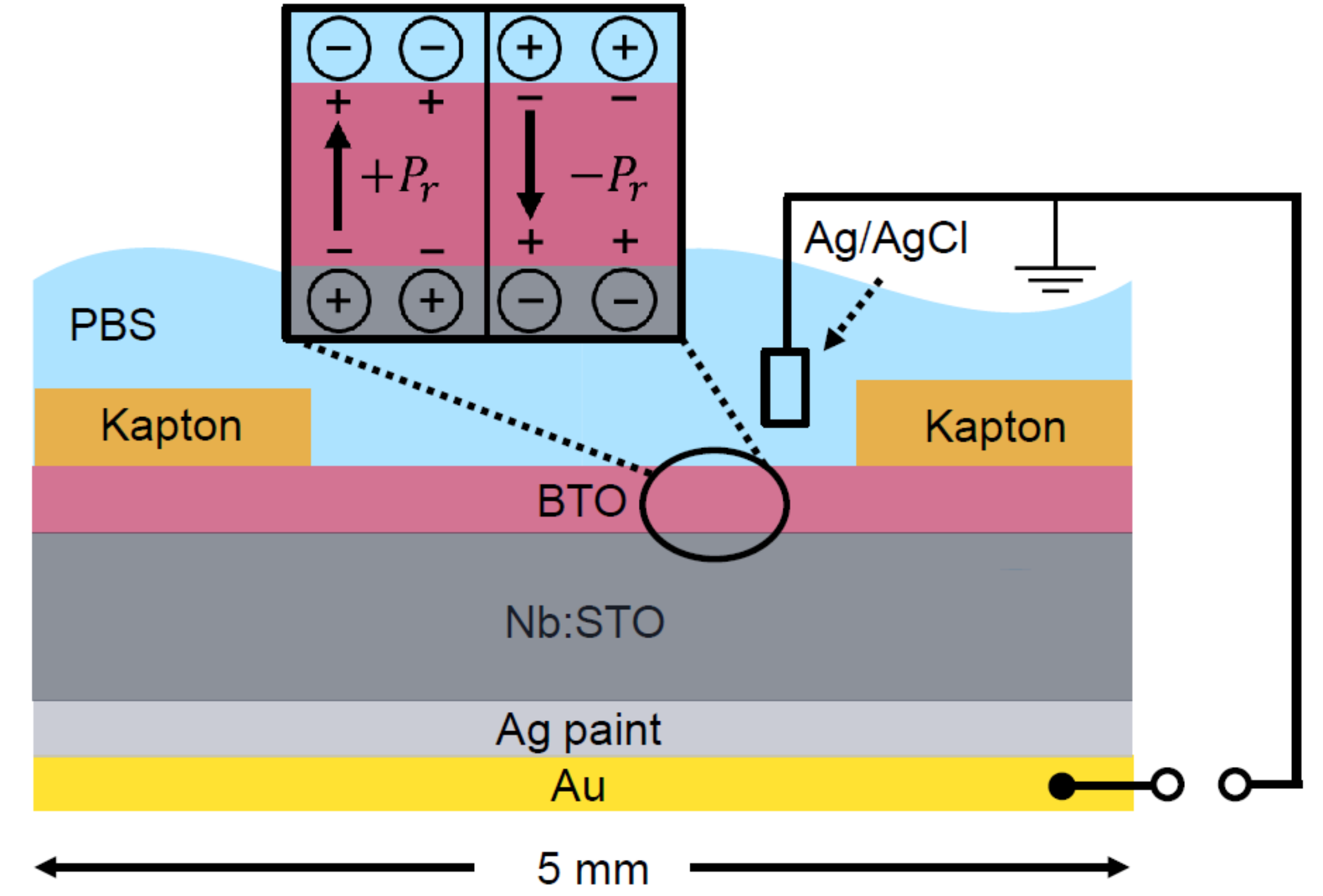}
\caption{Schematic sketch of the experimental set-up utilized for electrical measurements in EFS device configuration.
Kapton tape ($\varepsilon_{r}$ $\sim$ 3.5) is deposited at the edges of the BTO film, to define an active area of 13.7 mm$^2$ of the BTO surface, which is in direct contact with the liquid electrolyte.
The electrolyte consists of PBS and is connected with a grounded Ag/AgCl reference electrode.
}
\label{fig:Fig1} 
\end{figure} 
Figure~\ref{fig:Fig1} shows a schematic sketch of the experimental set-up utilized for electrical measurements in EFS device configuration.
Insulating Kapton tape with a low dielectric constant $\varepsilon_{r}$ $\sim$ 3.5 \cite{Simpson1997} was deposited on top of the films to define an active surface area (13.7 mm$^2$) of BTO, which is in direct contact with PBS acting as liquid electrolyte top contact.
The back of the EFS chips, were covered with conductive silver paint to ensure an Ohmic contact to a subsequently deposited Au layer which is connected to external electronics via Au conductor lanes insulated with polyimid.
We utilized the set-up depicted in Fig.~\ref{fig:Fig1} for cyclic voltammetry (CV) and electrochemical impedance spectroscopy (EIS) measurements on the EFS devices with a potentiostat in three-electrode configuration.
We used a Pt mesh counter electrode, an Ag/AgCl reference electrode and the EFS device acting as working electrode.
For all electrical measurements in EFS device configuration, the voltage was applied to the bottom contact and the Ag/AgCl reference electrode in the electrolyte was grounded.

\section{Results and Discussion} 
\label{sec:Results}
In the following, we present results from electrical measurements on bare BTO thin films and BTO films in EFS device configuration.
\subsection{Thin film characterization}
\label{subsec:Films}
\begin{figure*}[t] 
\includegraphics[width=1\textwidth]{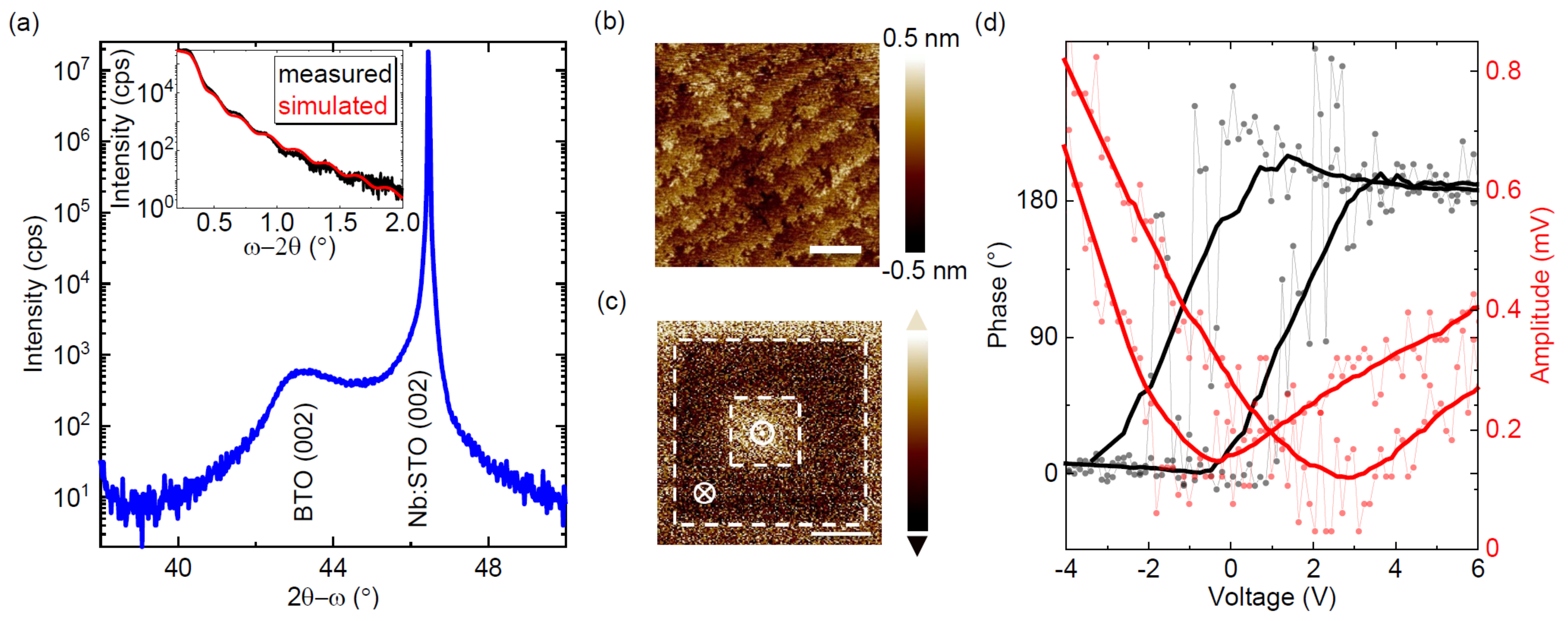}
\caption{Structural and electrical characterization of a bare BTO thin film deposited on (001)-oriented Nb:STO substrate. (a) XRD 2$\theta$-$\omega$ scan with XRR $\omega$-2$\theta$ scan in the inset (black) and its simulation (red). (b) AFM image of topography. Scale bar is 1~$\mu$m. (c) PFM out-of-plane response collected with V$_{\text{ac}}$ of 1~V after poling with V$_{\text{dc}}$ (-4V,~+6)~V. Scale bar is 1~$\mu$m. (d) PFM out-of-plane hysteresis loop collected with V$_{\text{ac}}$ of 1~V. Phase (black) and amplitude (red).}
\label{Fig2} 
\end{figure*} 

The high-resolution XRD scan along the 2$\theta$-$\omega$ axes of the BTO film is shown in Fig.~\ref{Fig2}(a). The diffraction peak at $\sim$\,43.1$^{\circ}$ corresponds to the (001)-oriented BTO film. The thickness of the BTO film is determined to be 18 nm from the simulations of XRR measurements using the Panalytical X'Pert Reflectivity Software, as depicted in the inset of Fig.~\ref{Fig2}(a). The surface roughness of the films was determined to be 0.6 nm by XRR.
The low surface roughness is corroborated by AFM measurements, see image in Fig.~\ref{Fig2}(b), which yields a root mean square roughness of 0.2 nm.
We carried out PFM measurements to check whether the BTO film is ferroelectric. We locally applied V$_{\text{dc}}$ to the sample while grounding the PFM tip and then detected the PFM response using V$_{\text{ac}}$ in the same configuration. In Fig.~\ref{Fig2}(c), the observation of two contrast levels in the PFM out-of-plane response after the application of an electric field confirms the presence of reversibly switchable polarization - ferroelectricity. 
We additionally performed switching spectroscopy using PFM out-of-plane, see Fig.~\ref{Fig2}(d). A typical ferroelectric response in amplitude and phase is observed for the V$_{\text{dc}}$ (-4,+6)\,V using a read voltage V$_{\text{ac}}$ of 1\,V.
Thus, we conclude that our BTO films are of high structural quality and ferroelectric.
\subsection{Device characterization}
\label{subsec:Films}
\begin{figure}[b] 
\includegraphics[width=1\columnwidth]{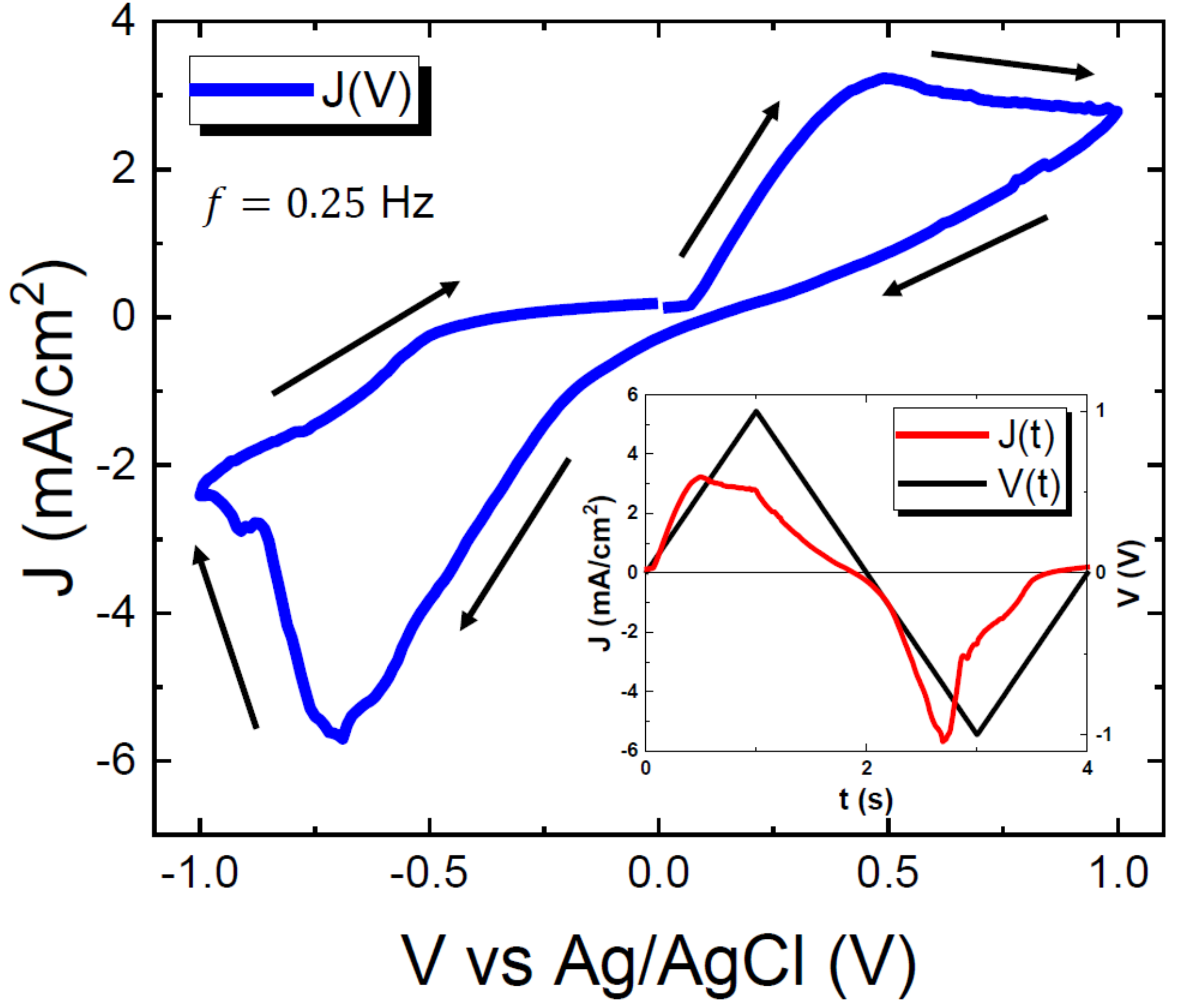}
\caption{Bipolar current-voltage ($J$-$V$) characteristics of the EFS device for a voltage scan $V(t)$ from 0 $\rightarrow$ 1 $\rightarrow$ -1 $\rightarrow$ 0 V at a sweep rate of 1~V/s. The inset shows the measured current density $J$ as a function of time $t$.
}
\label{fig:Fig3} 
\end{figure} 

Figure~\ref{fig:Fig3} shows the bipolar current-voltage ($J$-$V$) characteristics of the EFS device obtained from CV measurements for a voltage sweep $V(t)$ from 0 $\rightarrow$ 1 $\rightarrow$ -1 $\rightarrow$ 0 V, applied to the bottom electrode at a sweep rate of 1~V/s.
Prior to measurement, an identical voltage sweep was carried out on the device to align the ferroelectric polarization of the BTO layer downwards, which we refer to the state $P^{{\rm (FE)}}$ = -$\text{P}_{\text{r}}$.
In ferroelectrics, the relationship between displacement field and applied voltage is nonlinear and hysteretic (ferroelectric hysteresis loop), i.e. 
\begin{eqnarray} 
D(V(t))&=&C_sV(t)+P^{{\rm (FE)}}(V(t)) \quad.
\label{eq:D_FE}
\end{eqnarray} 
For a perfectly insulating ferroelectric material, conduction current $J_{\mathrm{c}}$ is absent and the $J$--$V$ loop contains only displacement current density $\partial D / \partial t$, i.e.
\begin{eqnarray} 
J_{\mathrm D}^{{\rm (FE)}}(V(t))&=&C_s \frac{\partial}{\partial t}V(t)+\frac{\partial}{\partial t} P^{{\rm (FE)}}(V(t))  \quad.
\label{eq:J_FE}
\end{eqnarray} 
Note, for common extracellular capacitive stimulation, ferroelectric polarization current $\partial P^{{\rm (FE)}}/\partial t $ is absent and so from Eq.~(\ref{eq:J_FE}), the current contribution is only $C_s \partial V(t)/\partial t$.
If the applied voltage signal exceeds the coercive voltage $\pm$ $V_{\mathrm{C}}$ of the ferroelectric, the displacement current Eq.~(\ref{eq:J_FE}) exhibits characteristic switching peaks at $\pm$ $V_{\mathrm{C}}$ due to the change in ferroelectric polarization $\partial P^{{\rm (FE)}}(V(t))/\partial t $.
In the bipolar $J$-$V$ loop shown in Fig.~\ref{fig:Fig3}, we observe peaks in the measured current density at +$V_{\mathrm{C}}$ $\sim$ 0.5 V and -$V_{\mathrm{C}}$ $\sim$ -0.7 V, which indicate ferroelectric switching.
The difference in the coercive voltages can be attributed to an internal bias field due to the asymmetric electrode configuration of the EFS device~\cite{Max20}.
Note, that the values of the coercive voltages in EFS device configuration are below the values of coercive voltages obtained by PFM [cf. Fig.~\ref{fig:Fig2}], which can be attributed to the frequency-dependence of the coercive field in ferroelectrics \cite{Scott1996}.
The $J$-$t$ plot of the measured current density is shown in the inset of Fig.~\ref{fig:Fig3}, which also allows for better comparison with extracellular currents utilized for neuronal stimulation experiments, where it is common to plot the stimulation current as a function of time $t$ \cite{Corna21}.

However, the $J$-$V$ loop also indicates the presence of conduction (leakage) currents  $J_{\mathrm{c}}$, which are pronounced in the segments 1 $\rightarrow$ 0 V and -1 $\rightarrow$ 0 V.
Due to the presence of conduction currents, it is not possible to extract the ferroelectric hysteresis loop Eq.~(\ref{eq:D_FE}) of the EFS device by integrating the bipolar $J$-$V$ loop depicted in Fig.~\ref{fig:Fig3}.
In general, a conduction current across an EFS structure must be related with an electrochemical reaction at the ferroelectric/electrolyte interface, i.e., a reducing reaction with electrons or an oxidizing reaction with holes in combination with a redox system in the electrolye with a suitable level of the redox potential \cite{Gerischer1990,Wallrapp06}.
It should be noted, that conduction currents are undesired for neural stimulation with ferroelectric microelectrodes, since they can affect safety and long-term stability \cite{Becker21,Oldroyd21}.
For BTO, it is known that oxygen vacancies $O_{o}$ can act as donors of free electrons $e^{-}$ which induce $n$-type conductivity \cite{Harris1978}. This process is described by the reaction
\begin{eqnarray} 
O_{o}&\rightarrow&\frac{1}{2}O_{2}+V_{o}^{2+}+2e^{-}  \quad,
\label{eq:Oo}
\end{eqnarray} 
where $V_{o}^{2+}$ denotes a doubly ionized oxygen vacancy.
The concentration of defects and associated charge carriers in BTO films is also dependent on the strain state of the films \cite{Aidoud17}.
In that context, we note that the conductivity in BTO is expected to be due to purely electronic processes, i.e., oxygen or cation vacancies are epitaxially clamped and immobile \cite{Tyunina21}.
In addition to the presence of free electrons in BTO, electrons may be injected across the Nb:STO/BTO interface of the EFS device, since the Nb:STO substrate is a heavily doped $n$-type semiconductor, which could form an Ohmic contact to electron-doped BTO.
It is important to note, that if the BTO exhibts properties of an $n$-type semiconductor, a space charge region (depletion layer) may exist at the BTO/electrolyte interface which dominates the resistive and capacitive properties of the EFS device \cite{Gerischer1990}.
The impact of a space charge region at the BTO/electrolyte interface on the electrical properties of the EFS device will be discussed in more detail below.
\begin{figure}[t] 
\includegraphics[width=1\columnwidth]{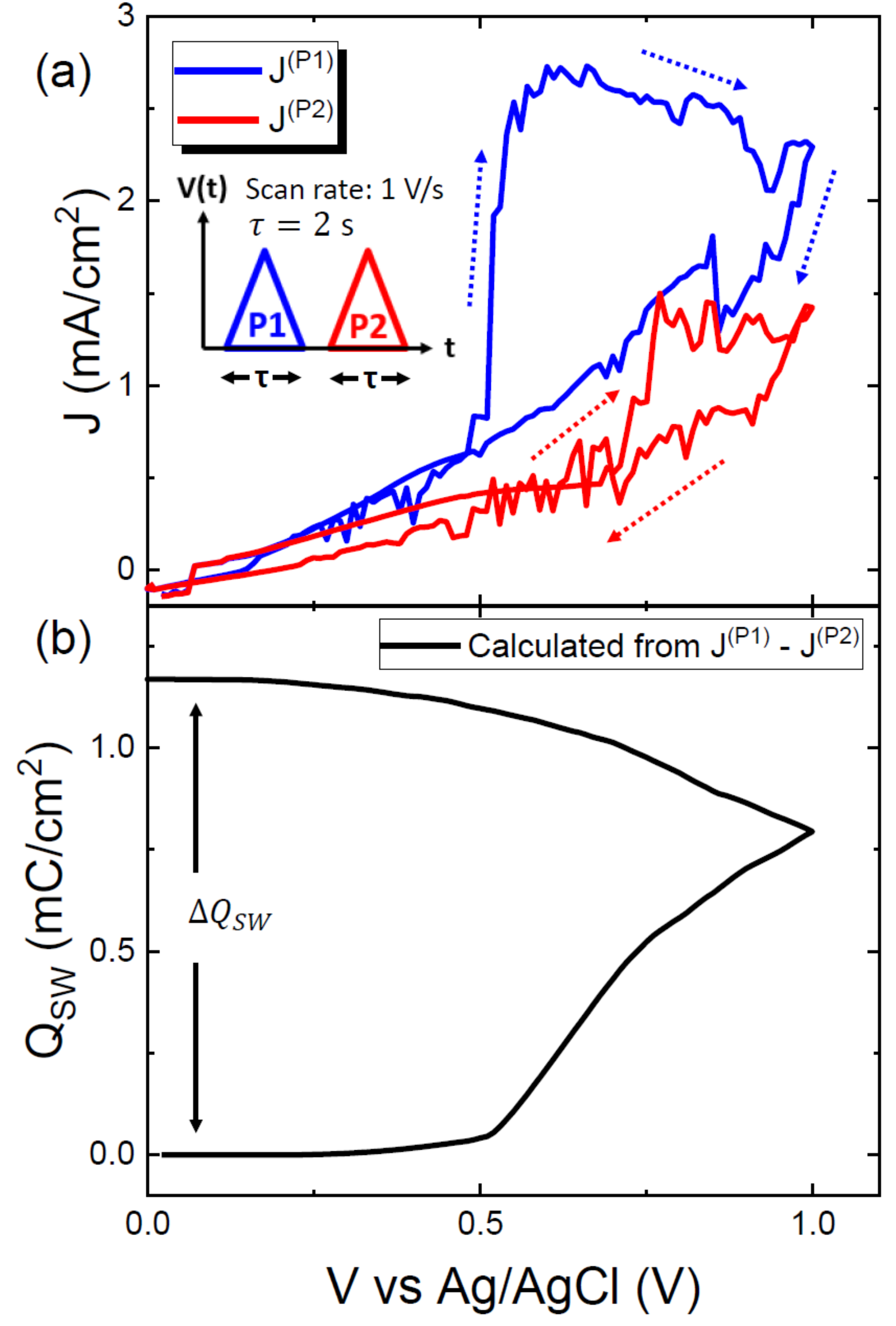}
\caption{(a) Current densities $J^{{\rm (P1)}}(V)$ and $J^{{\rm (P2)}}(V)$ of the EFS device in response to the unipolar voltage signals depicted in the inset, i.e., a first positive ($P1$) signal followed by a second positive ($P2$) signal. Prior to the $P1$ and $P2$ voltage signals, a negative voltage signal was applied to set the EFS device to the state -$\text{P}_{\text{r}}$.
(b) Nonlinear and hysteretic charge density loop $Q_{\rm SW}$ vs $V$ of the EFS device, calculated from the difference in the measured current responses depicted in (a).
}
\label{fig:Fig4} 
\end{figure} 

In order to confirm the ferroelectric nature of the observed current peaks in Fig.~\ref{fig:Fig3}, we investigated the current response of the EFS device to unipolar voltage signals.
The results are depicted in Fig.~\ref{fig:Fig4}.
Here, Fig.~\ref{fig:Fig4}(a) shows the current densities $J^{{\rm (P1)}}(V)$ and $J^{{\rm (P2)}}(V)$  of the EFS device in response to subsequently applied unipolar voltage signals of duration $\tau$ (see inset), i.e., a first positive ($P1$) signal followed by a second positive ($P2$) signal.
Prior to the $P1$ and $P2$ voltage signals, a negative voltage signal was applied to set the EFS device to the polarization state -$\text{P}_{\text{r}}$.
If ferroelectricity is present in the EFS device, the $P1$ signal induces ferroelectric polarization reversal from -$\text{P}_{\text{r}}$ to +$\text{P}_{\text{r}}$ and the current response contains ferroelectric switching current according to Eq.~(\ref{eq:J_FE}), whereas the $P2$ signal induces a current response, which contains no ferroelectric switching current.
From Fig.~\ref{fig:Fig4}(a), we clearly observe a difference between the current responses $J^{{\rm (P1)}}(V)$ and $J^{{\rm (P2)}}(V)$ of the EFS device, i.e., $J^{{\rm (P1)}}(V)$ exhibits a steep rise in the current at $\sim$ 0.5 V, which is absent in $J^{{\rm (P2)}}(V)$.
The observed difference between $J^{{\rm (P1)}}(V)$ and $J^{{\rm (P2)}}(V)$ is a fingerprint of ferroelectricity and rules out that electrochemical reactions (Faradaic processes) at the BTO/PBS interface are the origin of the observed current peaks in Fig.~\ref{fig:Fig3}.
Nonetheless, also the $J^{{\rm (P2)}}(V)$ is slightly hysteretic, which might be due to the backswitching of ferroelectric domains between the $P1$ and $P2$ voltage signals, caused by the existence of depolarization field due to imperfect screening of polarization charges \cite{Max20} or due to unsaturated polarization switching during the $P1$ pulse.
 Moreover, it should be noted, that also the conduction current response could be different for the voltage signals $P1$ and $P2$ (resistive switching behavior), which will be discussed in more detail below.

The charge density $Q_{\rm SW}$, which flows in response to polarization switching is calculated by integrating the difference in current response $\int_{0}^{\tau}[J^{(P1)}-J^{(P2)}]dt$ and the results are depicted in Fig.~\ref{fig:Fig4}(b).
Here, the $Q_{\rm SW}$--$V$ loop shows the typical nonlinear and hysteretic behavior, which is characteristic for ferroelectrics.
Thus, if conduction current is not present or equal for the two polarization states $\pm \text{P}_{\text{r}}$ of the EFS device (no resistive switching behavior), the total charge density $\Delta Q_{\rm SW}$ measured during polarization reversal from -$\text{P}_{\text{r}}$ to +$\text{P}_{\text{r}}$ would be equal to 2$\text{P}_{\text{r}}$ [cf. Eq.~(\ref{eq:CIC_FE})].
%
%
However, the value of the total charge density $\Delta Q_{\rm SW}$ $\approx$ 1.2 mC/cm$^2$ [cf. Fig.~\ref{fig:Fig4}(b)] corresponds to $\text{P}_{\text{r}}$ = 0.6 mC/cm$^2$ of the BTO thin film -- a higher value than typical polarization values of epitaxial BTO films, which are in the range below 0.1 mC/cm$^2$ \cite{Wu20}.
We identified two possible causes for this observation.
First, the duration $\tau$ of the utilized voltage signals corresponds to a low-frequency (0.5 Hz) measurement of polarization, which can result in an overestimation of the remanent polarization as compared to polarization measurements in the kHz regime  \cite{Fina11}. However, due to limitations in the scan-rate of the utilized potentiostat, we are unable to conduct measurements at significant higher frequencies for comparison.
Another possible explanation could be resistive switching behavior, i.e the conduction current $J_{\mathrm{c}}$ is different for the two different polarization states of the EFS device. As a consequence, the contribution from $J_{\mathrm{c}}$ is not fully eliminated in the $Q_{\rm SW}$--$V$ loop, which is obtained by integrating the difference of the current densities $J^{{\rm (P1)}}(V)$ and $J^{{\rm (P2)}}(V)$.
Here, the slightly higher current of $J^{{\rm (P1)}}(V)$ as compared to $J^{{\rm (P2)}}(V)$ in the range below 0.5 V [cf. Fig.~\ref{fig:Fig4}(a)] suggests that the EFS device has a lower resistance in the polarization state -$\text{P}_{\text{r}}$ as compared to to the state +$\text{P}_{\text{r}}$.

To investigate the resistive switching behavior of the EFS device in more detail, we conducted EIS measurements.
\begin{figure}[t] 
\includegraphics[width=1\columnwidth]{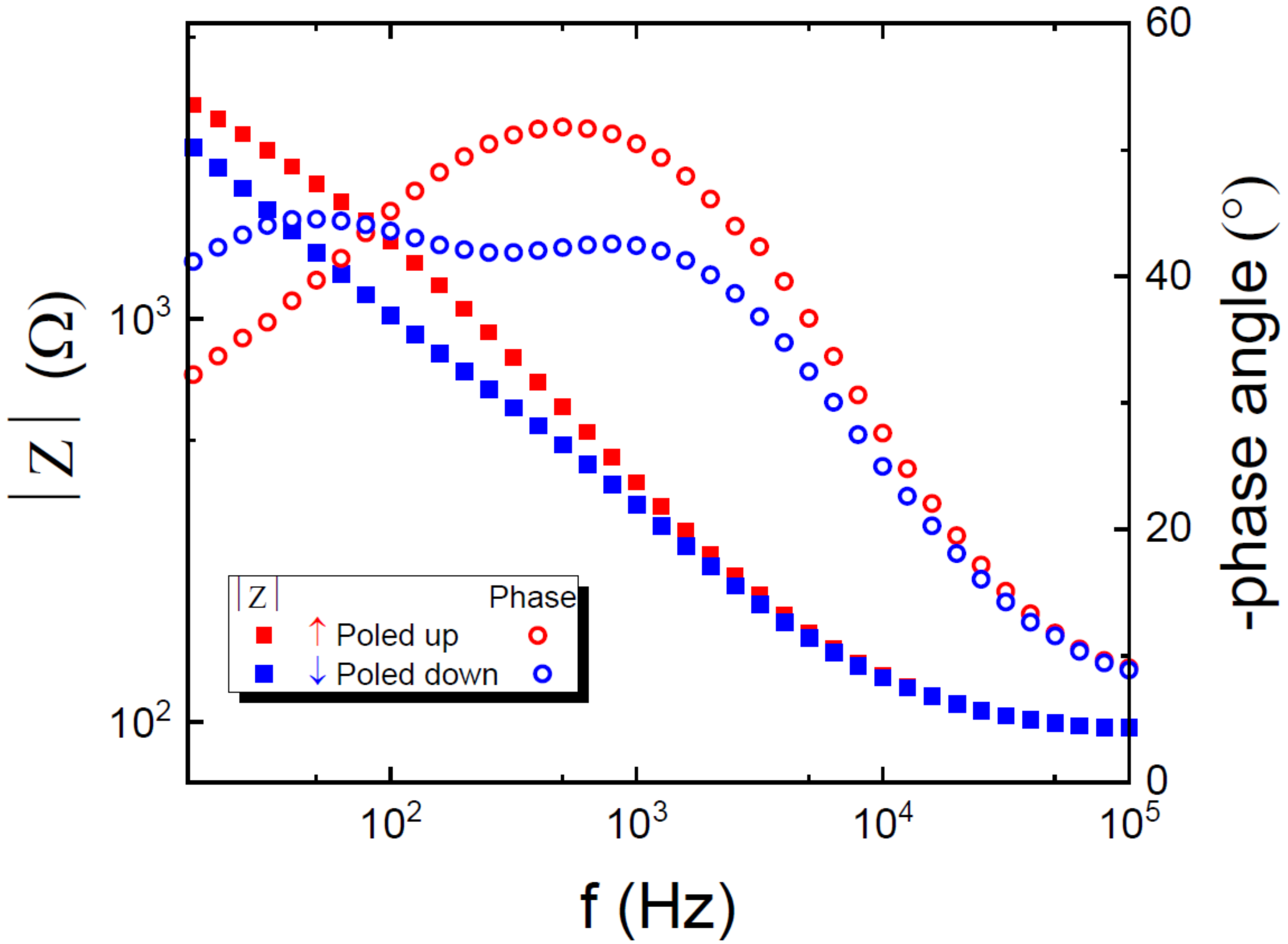}
\caption{Impedance modulus $\vert$Z$\vert$ and phase angle obtained by EIS measurements on the EFS device in the state +$\text{P}_{\text{r}}$ (red symbols)  and  -$\text{P}_{\text{r}}$ (blue symbols).
}
\label{fig:Fig5} 
\end{figure} 
The results are shown in Fig.~\ref{fig:Fig5}, where the measured impedance modulus $\vert$Z$\vert$ and phase angle is plotted as a function of frequency $f$ for the polarization states +$\text{P}_{\text{r}}$ and -$\text{P}_{\text{r}}$.
We clearly identify a difference in the impedance response for the two polarization states $\pm \text{P}_{\text{r}}$ of the EFS device which demonstrates resisitive switching behavior.
Here, the difference is more pronounced at frequencies below $\sim$~1 kHz, whereas at higher frequencies, the difference vanishes.
This is because at high frequencies, the serial resistance of the electrolyte dominates the impedance \cite{BardFaulkner01}.

In order to extract quantitative values for the specific resistance and the specific capacitance for the two polarization states $\pm \text{P}_{\text{r}}$ of the EFS device, we evaluated the measured impedance response with an equivalent-circuit consisting of an effective capacitor $C_{\rm eff}$ in parallel with an effective resistor $R_{\rm eff}$ \cite{Brown05}.
For the impedance evaluation, we chose a frequency of 100 Hz, to minimize the influence of the serial resistance of the electrolyte and for better comparison of our results with earlier work on dielectric TiO$_2$ an HfO$_2$ in electrolyte--oxide--silicon (EOS) configuration for capacitve stimulation applications \cite{Wallrapp06}. 
For the state +$\text{P}_{\text{r}}$, our analysis yields $\rho_{\rm eff}$ = 16.8 k$\Omega$cm and an effective specific capacitance of $C_{s,\rm eff}$ = 5.3 $\mu$F/cm$^2$, which corresponds to an effective relative dielectric contant of 108.
Polarization reversal in the BTO layer sets the EFS device to the state -$\text{P}_{\text{r}}$, which is characterized by $\rho_{\rm eff}$ = 10.7 k$\Omega$cm and  $C_{s,\rm eff}$ = 7.8 $\mu$F/cm$^2$ (effective dielectric constant of 159).
Thus, the EFS device exhibts resistive switching behavior and the state -$\text{P}_{\text{r}}$ corresponds to the low resistance state of the EFS device, where the specific resistance is reduced by 36 \% as compared to the state +$\text{P}_{\text{r}}$.

We will now discuss the physical mechanism of the resistive switching behavior in the EFS device in more detail.
So far, a comprehensive theory of polarization-modulated charge transport across a (semiconducting) ferroelectric/electrolye interface is missing.
However, a model for the influence of ferroelectric polarization on the properties of a (semiconducting) ferroelectric/metal Schottky contact exists \cite{Blom1994,Pintilie05_I}.
In case of a ferroelectric/metal Schottky contact, the depletion width of the space charge region inside the ferroelectric is given by \cite{Pintilie05_I}
\begin{eqnarray} 
w&=&\sqrt{\left|\frac{\varepsilon_{0}\varepsilon_{{\rm st}}(V+V_{{\rm bi}}')}{eN_{{\rm eff}}}\right|} \quad,
\label{eq:w}
\end{eqnarray} 
where $\varepsilon_{{\rm st}}$ is the static dielectric constant of the depleted region, $e$ is the electron charge, $N_{{\rm eff}}$ is the effective charge density in the depleted region and $V_{{\rm bi}}'$ is the apparent built-in potential which is modified by the ferroelectric polarization according to \cite{Pintilie05_I}
\begin{eqnarray} 
V_{{\rm bi}}'&=&V_{{\rm bi}}\pm P_{{\rm r}}\frac{\delta}{\varepsilon_{0}\varepsilon_{{\rm st}}} \quad,
\label{eq:vbi}
\end{eqnarray} 
with $\delta$ being the distance between the surface polarization charge and the interface of the Schottky contact.
According to Eqs.~(\ref{eq:w}) and~(\ref{eq:vbi}), the negative polarization state -$\text{P}_{\text{r}}$ lowers the apparent built-in potential $V_{{\rm bi}}'$ and decreases the depletion width $w$ of the Schottky contact.
As a consequence, the Schottky contact lowers its resistance and increases its capacitance, which is similar to the behavior of the EFS device.
The bipolar $J$--$V$ analysis of the EFS device [cf. Fig.~\ref{fig:Fig3}], already suggested that the BTO layer exhibits $n$-type semiconducting properties due to the presence of free electrons.
Thus, we conclude that a space charge region (depletion layer) exists inside the BTO layer adjacent to the electrolyte, which dominates the impedance response of the EFS device \cite{Harris1978,Gerischer1990}.
In analogy to the model of ferroelectric/metal interface \cite{Pintilie05_I}, the space charge region in the BTO layer is modulated by the ferroelectric polarization of the semiconducting BTO layer and hence causes the observed ferroelectric resistive switching behavior.

For neural stimulation with ferroelectric microelectrodes, resistive switching behavior is undesired since electrochemical charge transport (conduction current $J_{\mathrm{c}}$) across the ferroelectric/electrolyte interface may negatively affect long-term stability \cite{Becker21, Oldroyd21}.
However, resistive switching behavior of two-terminal EFS devices could be exploited for extracellular recording in hybrid neurocomputational systems: a biological neuron can be cultivated on the active area of the EFS device and the extracellular voltage generated by an action potential of the neuron acts as driving voltage to change the resistance state of the EFS device.
For our EFS device, the coercive field of $\sim$ 0.5 V is too high to achieve polarization switching induced by a firing neuron.
However, very recently, ultra-low ferroelectric switching voltages in the order of $\sim$~0.1 V have been reported for BaTiO$_3$ films \cite{Jiang22}, which is in the range of the extracellular voltage generated by a neuronal action potential.
In addition, we note that there is every reason to believe that also a continous change in resistance (memrisitve behavior) could be achieved in the EFS device by altering the ferroelectric domain structure with subcoercive voltage pulses \cite{Chanthbouala12}. 
The impedance response of memristive EFS devices in intermediated resistance states can be investigated by domain wall pinning element modeling \cite{BeckerJAP22}, which may help to improve device performance.
EFS memristors -- acting as artificial synapses -- with ultra-low switching voltages may be utilized in the future for direct coupling to a biological neuron, resulting in the formation of a two-terminal neuron-ferroelectric junction.
Such a neuron-ferroelectric junction is in contrast to recent experiments, where a coupling between an the signals of an indidual MEA chip and an individual memristor has been achieved \cite{Serb20}.

\section{Conclusion} 
\label{sec:Conclusion}
In conclusion, polarization reversal in epitaxial 18-nm-thick ferroelectric BTO films with phosphate-buffered saline (PBS) acting as liquid electrolyte top contact is demonstrated.
This is an important step for the realization of bioelectronic devices based on ferroelectric microelectrodes.
In addition, the investigated electrolyte-ferroelectric-semiconductor (EFS) device exhibits polarization-modulated resisitve switching behavior.
This effect could be exploited in future in hybrid neurocomputational systems combining artificial and biological neurons.

\section*{Acknowledgements} 
\label{sec:ACK}
We acknowledge funding from the ERC grant EU-H2020-ERC-ADG \# 882929 EROS, the Royal Academy of Engineering Chair in Emerging Technologies grant - CIET1819\_24, the Nano Doctoral training centre EPSRC grant EP/S022953/1, and the Swiss National Science Foundation grant P2EZP2-199913.
%



\newpage

\bibliography{FE-Microelectrode}
\end{document}